\documentclass[12pt]{article}
\usepackage{graphicx}
\usepackage{amssymb}
\def\beq{\begin{equation}}
\def\eeq{\end{equation}}
\def\bea{\begin{eqnarray}}
\def\eea{\end{eqnarray}}

\def\roughly#1{\mathrel{\raise.3ex\hbox
{$#1$\kern-.75em\lower1ex\hbox{$\sim$}}}}

          \def\s{\tau}
     \def\t{\tilde}        
  \def\rr2{{1\over\sqrt{2}}}

\def\.{\!\cdot\!}    \def\:{\cdots}   \def\[{\left[}   \def\]{\right]}
\def\({\left(} \def\){\right)} 
\def\t3{{1\over\sqrt{3}}}
\def\s6{{1\over\sqrt{6}}}

\begin{document}
\begin{flushright}
UMISS-HEP-2011-05 \\
[10mm]
\end{flushright}

\begin{center}
\bigskip{\Large  \bf Deviation from Tri-Bimaximal Mixing and Large Reactor Mixing Angle.}  
\\[8mm]
Ahmed Rashed 
\footnote{E-mail:
\texttt{amrashed@phy.olemiss.edu}}
\end{center}

\begin{center}
{\it  Department  of Physics and Astronomy,}\\ 
{ \it University of Mississippi,}\\
{\it  Lewis Hall, University, MS, 38677, USA}\\
\end{center}

\begin{center}
{\it  Department  of Physics, Faculty of Science,}\\
{\it  Ain Shams University, Cairo, 11566, Egypt.}\\
\end{center}

\begin{center} 
\bigskip (\today) \vskip0.7cm {\Large Abstract\\} \vskip10truemm
\parbox[t]{\textwidth}
{Recent observations for a non-zero $\theta_{13}$ have come from various experiments. We study a model of lepton mixing with a $2-3$ flavor symmetry to accommodate the sizable $\theta_{13}$ measurement. In this work, we derive deviations from the tri-bimaximal (TBM) pattern arising from breaking the flavor symmetry in the neutrino sector, while the charged leptons contribution has been discussed in a previous work.  Contributions from both sectors towards accommodating the non-zero $\theta_{13}$ measurement are presented.}

\end{center}

\thispagestyle{empty} \newpage \setcounter{page}{1}
\baselineskip=14pt


\section{Introduction}

Neutrino oscillations can be parametrized in terms of three mixing angles $\theta_{12},\;\theta_{13},\;\theta_{23}$ and Dirac $(\delta)$ and Majorana $(\zeta_1,\; \zeta_2)$ CP violating phases
\begin{eqnarray}
V = \left( \begin{array}{ccc} c^{}_{12} c^{}_{13} & s^{}_{12}
c^{}_{13} & s^{}_{13} e^{-i\delta} \\ -s^{}_{12} c^{}_{23} -
c^{}_{12} s^{}_{13} s^{}_{23} e^{i\delta} & c^{}_{12} c^{}_{23} -
s^{}_{12} s^{}_{13} s^{}_{23} e^{i\delta} & c^{}_{13} s^{}_{23} \\
s^{}_{12} s^{}_{23} - c^{}_{12} s^{}_{13} c^{}_{23} e^{i\delta} &
-c^{}_{12} s^{}_{23} - s^{}_{12} s^{}_{13} c^{}_{23} e^{i\delta} &
c^{}_{13} c^{}_{23} \end{array} \right) P^{}_\nu \;
,\label{standpara}
\end{eqnarray}
where $c^{}_{ij}\equiv \cos \theta^{}_{ij}$, $s^{}_{ij} \equiv \sin
\theta^{}_{ij}$, and $P_\nu^{} \equiv \{1, e^{i\zeta_1}, e^{i\zeta_2}\}$ is
a diagonal phase matrix, which is physically relevant if neutrinos are Majorana particles. 
The experiments have shown that the angles of lepton mixing are relatively larger than their counterparts in the quark sector. Recent data from the Double Chooz \cite{Double-Chooz}, Daya Bay \cite{Daya-Bay},
RENO \cite{RENO} experiments as well as latest ${\rm T2K}$ \cite{Abe:2011sj} and MINOS \cite{MINOS} experiments have yielded nonzero values for $\theta_{13}$. The best-fit values for the mixing angles are given as \cite{Tortola:2012te}
\begin{eqnarray}
\sin^2 \theta_{12} &=& 0.320, \nonumber\\
\sin^2 \theta_{23}&=& 0.427 \; (0.600)\;\; \mbox{(for normal (inverted) hierarchy)},\nonumber\\
\sin^2 \theta_{13}&=& 0.0246 \; (0.0250)\;\; \mbox{(for normal (inverted) hierarchy)}.
\label{globalfit}
\end{eqnarray}

The distribution of the flavors in the mass eigenstates, corresponding to the best-fit values of the mixing angles, has shown that the leading order mixing method is a quite successful way to describe the lepton mixing. The most common patterns that have been discussed in the literatures to describe the lepton mixing, which may arise from discrete symmetries, are called; democratic (DC) \cite{minzhu}, bimaximal (BM) \cite{bi}, and tri-bimaximal (TBM) \cite{tri} mixing matrix. Recently, a pattern has been proposed to link the lepton and quark sectors so-called Tri-bimaximal-Cabibbo mixing \cite{King:2012vj}. 
%
Many previous studies have considered the TBM form in the symmetric limit of different flavor symmetries \cite{Lam, others-1, others-2,others-3,others-4,others-5,others-6,others-7,others-8,others-9}. Contributions from the charged lepton sector to the leptonic mixing have been studied previously \cite{CCC-1,CCC-2}. The recent $\theta_{13}$ measurement has been discussed \cite{recent-1,recent-2,recent-3,recent-4,recent-5}, several papers have considered deviations from the charged lepton sector \cite{T2K-cited-CL-NonDiag-1-1,T2K-cited-CL-NonDiag-1-2}. Early studies of a sizable $\theta_{13}$ have been conducted previously \cite{early}.

The leptonic mixing matrix is obtained from the contributions of the diagonalization of the charged lepton and neutrino mass matrices. Many models have been introduced to study the leptonic mixing in the basis where the charged lepton mass matrix is diagonal. Our approach considers both contributions from the charged lepton and neutrino sector to obtain the leading order leptonic mixing as well as deviations from it. One of the central ideas of this approach is the requirement that the mass matrices, in a symmetric limit, be diagonalized by unitary matrices composed of pure numbers  independent of the parameters of the mass matrices. If one starts with a $2-3$ symmetric mass matrix for the charged lepton sector and requires it to be diagonalized by unitary matrices of pure numbers one recovers the decoupled $2-3$ symmetry; decoupling of the first generation from the second and third generations. This helps in understanding the mass splitting between the first generation from the second and third generations.

Before we begin our analysis we would like to remark the fact that the quark and charged leptons exhibit similar hierarchical structures. We therefore assume the same flavor structure for them. One can use similar parametrization and flavor symmetric limit in the quark and charged lepton sector. The discussion of the 2-3 flavor symmetry in the quark sector can be found in Ref.~\cite{model23}. 


In Ref.~\cite{Rashed:2011zs}, they have considered the decoupled $2-3$ symmetry as the flavor symmetry in the charged lepton sector. The contributions of the charged lepton and neutrino sector have been discussed in Ref.~\cite{Rashed:2011zs} with the Bimaximal (BM) pattern being the leading order term of the lepton mixing. In this work, we assume a certain texture for the neutrino mass matrix with the third generation decoupled from the first two generations. Requiring the elements of the unitary matrix that diagonalizes the neutrino mass matrix to be independent of the mass parameters, the leptonic mixing turns out to have the TBM form in the symmetric limit under a certain condition. 

In our model, we introduce a Lagrangian that extends the SM particle content by three right-handed neutrinos, three complex singlet scalar fields, and an additional Higgs doublet.{\footnote{ Some recent motivations for considering two Higgs doublet models can be found in Ref.~\cite{2HDM}.}} The symmetry group of the SM is extended by the product of the symmetries $Z_4\times U(1) \times SO(3)$. The $Z_4$ symmetry serves to have a $2-3$ symmetric Yukawa matrix in the charged lepton sector and yields the mass matrices in the charged lepton and neutrino sector to have decoupled structures. The global $SO(3)$ flavour symmetry leads to diagonal Dirac neutrino mass matrix and equal vacuum expectation values of the three singlet scalars. The $SO(3)$ symmetry is broken in the other terms of the Yukawa Lagrangian. We present a global $U(1)$ symmetry that equates certain couplings of the neutrinos as we relate the couplings to the $U(1)$ charges. The $U(1)$ symmetry forbids the Majorana masses of the right-handed neutrinos. The Majorana neutrino masses are generated via the v.e.v of the singlet scalars and the $U(1)$ gets broken spontaneously. Without altering the lepton mixing, an additional Majorana mass term is introduced to protect one of the neutrino masses from blowing up.

Breaking the symmetry in the charged lepton sector has been studied in Ref.~\cite{Rashed:2011zs}. In the neutrino sector we generate a deviation to the TBM mixing by explicitly breaking the $SO(3)$ symmetry in the scalar potential. The symmetry breaking term violates the alignment of the v.e.v's of the singlet scalar fields. The contribution of the neutrino sector to the deviation of the lepton mixing goes as $\sim \frac{v^2}{w^2}$ where $v$ is the electroweak (EW) scale and $w$ is the scale of the v.e.v of singlet scalars. Introducing a small symmetry breaking parameter is sufficient to generate the realistic lepton mixing and mass matrices. If the realistic neutrino mass matrix has small deviations from the TBM form, one can say that the TBM is not an accidental symmetry \cite{Abbas:2010jw}.

The paper is organized in the following manner: In Sec.~2 we study the TBM mixing in the flavor symmetric limit. In Sec.~3 we discuss the Lagrangian that describes the flavor symmetry in the charged lepton and neutrino sector. In Sec.~4 we break the flavor symmetry  to generate the realistic leptonic mixing. In Sec.~5 we show the numerical results due to the symmetry breaking, and, finally, in Sec.~6 we conclude with a summary of the results reported in this
work.


\section{The TBM matrix from flavor symmetry}


In Ref.~\cite{Rashed:2011zs}, it was assumed that the Yukawa matrix of the charged lepton sector is invariant under the $\mu-\tau$ interchange. Since the leptonic mixing matrix is composed of pure numbers, it is naturally supposed the mass matrices to be diagonalized by a unitary matrices composed of pure numbers. This results in the Yukawa matrix of the charged lepton sector to be decoupled as \cite{Rashed:2011zs}
\bea
Y^L_{23} & = &
\pmatrix{l_{11} & 0
&  0  \cr  0  &  \frac{1}{2}{l_{T}}  &  \frac{1}{2}{l_{T}}\cr  0  &\frac{1}{2}{l_{T}}  &\frac{1}{2}{l_{T}}}. \
 \label{23sym}
 \eea
This Yukawa matrix leads to zero muon mass $m_\mu =0$, shifting $m_\mu$ from the zero value will cause symmetry breaking later. The Yukawa matrix can be invariant under additional flavor symmetries such as a $Z_2$ symmetry 
and diagonalized by the unitary matrix $W^{l}_{23}$ given by
\bea
W^l_{23} &  = & \pmatrix{1 & 0
&   0  \cr   0   &  -\frac{1}{\sqrt{2}}   &  \frac{1}{\sqrt{2}}\cr   0
&\frac{1}{\sqrt{2}} &  \frac{1}{\sqrt{2}}}.\
\label{wl} 
\eea

Having established the flavor symmetry in the charged lepton sector, we use the TBM mixing as an input to identify the flavor symmetry in the neutrino sector. 
In the 2-3 symmetric limit the PMNS matrix, with $s_{13}=0$, is given by
\bea
U_{PMNS}^{s} & = & \pmatrix{c_{12} &s_{12}&0\cr
-\rr2 s_{12}
&\rr2 c_{12}  & \rr2\cr
\rr2 s_{12}
&- \rr2 c_{12} &
 \rr2}.
\eea
The TBM form is obtained by setting 
\bea
s_{12}&= & \sin{\theta_{12}}= { 1 \over {\sqrt{3}}}, \nonumber\\
c_{12}& = & \cos{\theta_{12}}= \sqrt{{2 \over 3}}.\
\label{solar}
\eea
Then, we can express $U_{PMNS}^{s}$ as
\bea
U_{PMNS}^{s}&= U^{\dagger}_\ell U_\nu, \
\label{nulepton}
\eea
where
\bea
U_\ell & = & W^l_{23}, \nonumber\\
U_\nu & = & \pmatrix{c_{12}&s_{12}&0\cr s_{12}&-c_{12}&0\cr 0&0&1\cr}.\
\eea 
Let us  discuss the structure of the neutrino matrix in the flavor symmetric limit. It can be easily seen that ${\cal {M}}_\nu$ is given as
\bea
{\cal {M}}_\nu & = &
\pmatrix{
a  & \sqrt{2}(a-b) & 0 \cr
\sqrt{2}(a-b)  & b & 0 \cr
0  & 0 & c
},\
\label{neu_mass}
\eea
with $U^T_\nu{\cal M}_\nu U_\nu={\cal M}_\nu^{d}$ and mass eigenvalues given by 
\beq
{\cal {M}}_\nu^{d}=diag\;(2a-b,\; 2b-a,\; c).
\label{mass-eigenvalues}
\eeq
We see that the neutrino mass matrix exhibits decoupling of the first two generations from the third one and it can be invariant under flavor symmetries such as a $Z_2$ symmetry.


\section{The origin of the $\mu-\tau$ symmetry in this model}


The Lagrangian that describes this model will be discussed in this section. It is assumed to be invariant under the product of the symmetries $Z_4 \times U(1)$. The Yukawa Lagrangian exhibits $\mu-\tau$ symmetry, which can be generated by a $Z_4$ symmetry. We use the see-saw mechanism to produce the neutrino mass matrix. The particle content of the model is given by
\begin{itemize}
 \item three left-handed lepton doublets $D_{\alpha_L}$, where $\alpha$ is denoted by $e,\; \mu,$ and $\tau$,
 \item three right-handed charged-lepton singlets $\alpha_R$, and
 \item three right-handed neutrino singlets $\nu_{\alpha R}$.
\end{itemize}
In the scalar sector, we employ
\begin{itemize}
 \item two Higgs doublets $\phi_j$ with vacuum expectation values, v.e.v's, $\left\langle 0|\phi_j^0|0\right\rangle =\frac{v_j}{\sqrt{2}}$, $j=1,2$,  and
 \item three complex singlet scalar fields $\epsilon_k$ with v.e.v's $\left\langle 0|\epsilon_k^0|0\right\rangle =w_k,\; \; k=1,2,3 $  .
\end{itemize}
The symmetry of the Lagrangians is assumed as
\begin{eqnarray}
 Z_4&:&D_{\mu_L}\leftrightarrow -D_{\tau_L},\; \mu_R\leftrightarrow -\tau_R,\;\nu_{\mu R}\leftrightarrow -\nu_{\tau R},\nonumber \\
&&  \nu_{e R} \rightarrow i \nu_{e R} ,\; e_R \rightarrow i e_R ,\; D_{e_L} \rightarrow i D_{e_L}, \nonumber \\
  && \epsilon_1 \rightarrow -i \epsilon_1 ,\;  \epsilon_2 \rightarrow i \epsilon_2 ,\; \epsilon_3 \rightarrow - \epsilon_3,\; \phi_1 \rightarrow \phi_1 ,\; \phi_2 \rightarrow \phi_2 , \nonumber\\
U(1)&:&  \left\{\nu_{(e,\mu,\tau) R},\;e_R ,\; (\mu,\tau)_R , \; D_{(e,\mu,\tau)_L},\;  \epsilon_{(1,2,3)} ,\; \phi_1 , \phi_2  \right\} =  \left\{ \frac{1}{3},\frac{7}{3},\frac{4}{3},\frac{4}{3} , \frac{2}{3}, 0,-1\right\}.\nonumber\\
\label{symmetry}
\end{eqnarray}
The most general Lagrangian invariant under the underlined symmetry is given by 
\bea
{\cal L}_Y & = & y_{1} \bar{D}_{e_L}e_R \phi_2 + \left[ y_2\left(\bar{D}_{\mu_L}\mu_R+\bar{D}_{\tau_L}\tau_R \right) +y_2\left(\bar{D}_{\mu_L}\tau_R+\bar{D}_{\tau_L}\mu_R \right)\right] \phi_1  \nonumber \\
 & +& y_D \left[ \bar{D}_{e_L} \nu_{eR}+  \bar{D}_{\mu_L} (\nu_{\mu R}+\nu_{\tau R})+\bar{D}_{\tau_L} (\nu_{\mu R}+\nu_{\tau R})  \right] \tilde{\phi_2}\nonumber \\
 & + &  \frac{1}{2}y \bar{\nu}_{e R} \left(\nu_{\mu R}^c \frac{(\epsilon_{1}+\epsilon_{2})}{\sqrt{2}}+\nu_{\tau R}^c \frac{(\epsilon_{1}-\epsilon_{2})}{\sqrt{2}} \right)  \nonumber \\ 
&+& \frac{1}{2}y \; \bar{\nu}_{e R}\; \nu_{e R}^c \;\epsilon_{3}+ h.c.
\label{Lagrangian-0}
\eea
Here, $\tilde{\phi}_j \equiv i\sigma_{2}\phi^*_j$ is the conjugate Higgs doublet. The $Z_4$ symmetry yields the decoupling structure in the charged lepton and neutrino mass matrices. In our model we relate the couplings to the $U(1)$ charges as $y =c q$ where $y$ is a coupling, $q$ is a $U(1)$ charge, and $c$ is a constant. This leads to a universal coupling to the right-handed neutrinos and to the charged leptons.

The phenomenology of the above Lagrangian with the off-diagonal elements $\bar{D}_{\mu_L} \nu_{\tau R}+\bar{D}_{\tau_L} \nu_{\mu R}$ can be studied. But in this model we choose to work with diagonal Dirac neutrino mass matrix $M_D$ to make the model even simpler. For this, we impose an approximate symmetry of the Lagrangian. A global $SO(3)$ flavour symmetry is introduced in a way that the transformations of the fields are given as follows:
\beq
\pmatrix{
e_R \cr \mu_R \cr \tau_R
},\;
\pmatrix{
D_{e_L} \cr D_{\mu_L} \cr D_{\tau_L}
},\;
\pmatrix{
\nu_{eR} \cr \nu_{\mu R} \cr \nu_{\tau R}
},\;
\pmatrix{
\epsilon_1 \cr \epsilon_2 \cr \epsilon_3
},\;
\phi_1 ,\; \phi_2 .
\eeq
In the above Lagrangian, the $SO(3)$ symmetry is only satisfied by the Dirac mass terms for the neutrinos and is broken by the other terms in the way that the Yukawa Lagrangian is invariant under the symmetry product $Z_4 \times U(1)$. The implications of proposing the $SO(3)$ flavour symmetry in the lepton sector will be discussed in a separate work. By implementing these particle assignment we find that the off-diagonal elements $\bar{D}_{\mu_L} \nu_{\tau R}+\bar{D}_{\tau_L} \nu_{\mu R}$ are forbidden leading to a diagonal Dirac mass matrix. 
We can then rewrite the above Lagrangian as
\bea
{\cal L}_Y & = & y_{1} \bar{D}_{e_L}e_R \phi_2 + \left[ y_2\left(\bar{D}_{\mu_L}\mu_R+\bar{D}_{\tau_L}\tau_R \right) +y_2\left(\bar{D}_{\mu_L}\tau_R+\bar{D}_{\tau_L}\mu_R \right)\right] \phi_1  \nonumber \\
 & +& y_D \left[ \bar{D}_{e_L} \nu_{eR}+  \bar{D}_{\mu_L} \nu_{\mu R}+\bar{D}_{\tau_L} \nu_{\tau R}  \right] \tilde{\phi_2}\nonumber \\
 & + &  \frac{1}{2}y \bar{\nu}_{e R} \left(\nu_{\mu R}^c \frac{(\epsilon_{1}+\epsilon_{2})}{\sqrt{2}}+\nu_{\tau R}^c \frac{(\epsilon_{1}-\epsilon_{2})}{\sqrt{2}} \right)  \nonumber \\ 
&+& \frac{1}{2}y \; \bar{\nu}_{e R}\; \nu_{e R}^c \;\epsilon_{3}+ h.c.
\label{Lagrangian}
\eea

When the singlet scalar fields acquire their v.e.v's, the $U(1)$ symmetry gets broken spontaneously and the neutrinos obtain their  Majorana masses \cite{U1-sym-breaking}. One of the neutrino masses blows up, therefore, we need to introduce a Majorana mass term as a $U(1)$ symmetry breaking term, which is not going to change the mixing, 
\begin{eqnarray}
 {\cal L}_M & = &\frac{1}{2} M \left[\bar{\nu}_{e R} \nu_{e R}^c  +\bar{\nu}_{\mu R} \nu_{\mu R}^c + \bar{\nu}_{\tau R} \nu_{\tau R}^c  \right] +h.c.
\label{Majorana}
\end{eqnarray}
The above Majorana mass term is invariant under the $SO(3)$ symmetry. 
The would-be-Goldstone bosons could be generated due to the spontaneous symmetry breaking of the global $U(1)$ symmetry by the v.e.v's of the singlet scalars. They can acquire masses through the explicit symmetry breaking of $U(1)$. Studying the effects of breaking the $U(1)$ symmetry falls beyond the main goal of this paper.

The most general scalar potential $V$ that is invariant under the above  symmetry product $Z_4 \times U(1) \times SO(3)$ is 
\bea
V &=& -\mu^2 \left( |\epsilon_1|^2 + |\epsilon_2|^2 + |\epsilon_3|^2 \right) + \left( |\epsilon_1|^2 + |\epsilon_2|^2 + |\epsilon_3|^2 \right) \sum_{i=1}^2 \sigma_i \phi_i^\dagger \phi_i   \nonumber\\
&& + \lambda \left( |\epsilon_1|^2 + |\epsilon_2|^2 + |\epsilon_3|^2 \right)^2 + V_{2HD}(\phi_1,\; \phi_2) ,
\label{potential}
\eea
where $V_{2HD}(\phi_1,\; \phi_2)$ is the potential of the two Higgs doublets \cite{thehiggshunter},
\bea
V_{2HD}(\phi_1,\; \phi_2) &=& \lambda'_{1} (\phi_1^\dagger \phi_1 -v_1^2)^2 + \lambda'_{2} (\phi_2^\dagger \phi_2-v_2^2)^2  \nonumber\\
 &+& \lambda'_{3} \left[(\phi_1^\dagger \phi_1 -v_1^2) + (\phi_2^\dagger \phi_2-v_2^2) \right]^2  \nonumber\\
 &+& \lambda'_{4} \left( (\phi_1^\dagger \phi_1 ) (\phi_2^\dagger \phi_2 ) - (\phi_1^\dagger \phi_2 ) (\phi_2^\dagger \phi_1 ) \right) 
\nonumber\\ 
&+&  \lambda'_{5}\left({\rm Re}(\phi_1^\dagger \phi_2 )-v_1 v_2 \cos \xi \right)^2  
+ \lambda'_{6}\left({\rm Im}(\phi_1^\dagger \phi_2 )-v_1 v_2 \sin \xi \right)^2, \nonumber\\
\eea
where $v_1$ and $v_2$ are the vacuum expectation values of the two Higgs doublets $\phi_1$ and $\phi_2$, consequently, and $\xi$ is a phase constant.
One can easily verify that the v.e.v's of the Higgs doublets are different and non-zero in the symmetric limit \cite{Rashed:2011zs}.

We can minimize the potential to get the v.e.v's  $(\left\langle 0|\epsilon_k^0|0\right\rangle =w_k )$ as follows
\bea
\left. \frac{\partial V}{\partial |\epsilon_1 |}\right|_{\mbox{min}}&=&-2\mu^2 w_1  +2w_1 \sum_{i=1}^2 \sigma_i v_i^\dagger v_i + 4\lambda w_1 \left( w_1^2 + w_2^2 + w_3^2 \right)=0, \nonumber\\
\left. \frac{\partial V}{\partial |\epsilon_2 |}\right|_{\mbox{min}}&=&-2\mu^2 w_2  +2w_2 \sum_{i=1}^2 \sigma_i v_i^\dagger v_i + 4\lambda w_2 \left( w_1^2 + w_2^2 + w_3^2 \right)=0, \nonumber\\
\left. \frac{\partial V}{\partial |\epsilon_3 |}\right|_{\mbox{min}}&=&-2\mu^2 w_3  +2w_3 \sum_{i=1}^2 \sigma_i v_i^\dagger v_i + 4\lambda w_3 \left( w_1^2 + w_2^2 + w_3^2 \right)=0. 
\eea
One can notice that the three equations are not independent. Thus, the three v.e.v's are the same and equal to
\beq
w^2 = \frac{\mu^2-(\sigma_1 |v_1|^2 + \sigma_2 |v_2|^2)}{6 \lambda},
\label{omega}
\eeq
where $w_k=w$ for $k=1,2,3$.

The explicit form of the charged lepton Yukawa matrix and the Majorana and Dirac neutrino mass matrices can be written from Lagrangian (\ref{Lagrangian}) as follows
\begin{eqnarray}
Y^L_{23} & = &\frac{v_1}{\sqrt{2}} \pmatrix{ y_1 v_2 /v_1   &  0  & 0   \cr  
                                              0   &  y_2  &  y_2   \cr  
                                              0   &  y_2  &  y_2   }, \nonumber\\
M_R & = & \pmatrix{M+\sqrt{2} v_w &  2 v_w &  0  \cr 2 v_w   &  M   &  0 \cr  0  & 0  & M  },\quad \mbox{with}\;\;  v_w=y\frac{w}{\sqrt{2}}, \nonumber\\
M_D & = & A~\textbf{I}, \quad \mbox{with}\;\; A= y_D \frac{v_2}{\sqrt{2}},
\label{PMNS-Yukawa}
\end{eqnarray}
where $\textbf{I}$ is the unit matrix. Using the see-saw formula \cite{seesaw}, the neutrino mass matrix is given as
\bea
{\cal M}_\nu & = & -M^T_D M^{-1}_R M_D.
\label{seesaw}
\eea
Then ${\cal M}_\nu$ has the structure 
\bea
{\cal M}_\nu & = & \pmatrix{X & G & 0 \cr G & Y & 0 \cr 0 & 0 & Z},
\label{mass-seesaw}
\eea
where 
\bea
X&=&-\frac{A^{2}M}{M^2 + \sqrt{2}M v_w - 4 v_w^2},\; Y=-\frac{A^{2} (M+\sqrt{2}v_w)}{M^2 + \sqrt{2}M v_w - 4 v_w^2}, \nonumber\\
G&=&\frac{2A^{2} v_w}{M^2 + \sqrt{2}M v_w - 4 v_w^2},\; Z=-\frac{A^{2}}{M }.
\label{numatrix2}
\eea
One can easily verify that the relation between the entries in Eq.~\ref{neu_mass} is satisfied by the elements in Eq.~\ref{numatrix2}. The mass eigenvalues $\left(2X-Y,2Y-X,Z\right)$ can be written as
\begin{eqnarray}
 m_1 & = & -\frac{A^2}{M +2\sqrt{2}v_w},\nonumber \\
 m_2 & = & -\frac{A^2}{M -\sqrt{2}v_w},\nonumber \\
 m_3 & = & -\frac{A^{2}}{M}.
\end{eqnarray}
From the above equations one can estimate the range of the v.e.v $v_2$ where $A= y v_2/\sqrt{2}$. As the absolute neutrino masses are in the eV scale, therefore, $v_2$ has to be in the MeV scale if the see-saw scale $(M)$ is in the TeV range. The mass eigenvalues satisfy the relation
\beq
\frac{1}{m_1}+\frac{2}{m_2}=\frac{3}{m_3}.
\eeq
Similar relations among the masses are discussed in Ref.~\cite{Barry:2010yk}. Thus, we can use the above sum-rule to obtain an upper limit for the heaviest mass $|m_3 |\leqslant \frac{3|m_1| |m_2|}{|2|m_1| + |m_2||}$ for the normal hierarchy or $|m_2 |\leqslant \frac{2|m_1| |m_3|}{|3|m_1| - |m_3||}$ for the inverted hierarchy.


\section{Symmetry Breaking}


The breaking of the flavor symmetries in the charged lepton and neutrino sector cause deviations from the TBM form. Symmetry breaking in the charged lepton sector has been considered in Ref.~\cite{Rashed:2011zs} to generate the realistic charged-lepton mass matrix
\bea
 Y^L &= &\pmatrix{l_{11} & l_{12} &  -l_{12} \cr l_{12} & \frac{1}{2}{l_{T}}(1+2\kappa_l) & \frac{1}{2}{l_{T} }\cr -l_{12}
&\frac{1}{2}{l_{T}} & \frac{1}{2}{l_{T}}},
\label{Yukawa-matrix-8}
\eea
with
\bea
l_{12} &\approx &\sqrt{\frac{z_\mu}{2}} (l_e - l_\mu), \nonumber\\
l_{T} & \approx & (l_{\tau}-l_{\mu}) (1-\frac{1}{2}(z z_\mu)^2),\nonumber\\
\kappa_l &= &z z_\mu , \nonumber\\
z_\mu &\equiv &{ m_{\mu} \over m_{\tau}}.
\eea
where $z$ is an arbitrary parameter with a value around 2. In this section we are going to consider deviations of the TBM structure from the neutrino sector.

We are going to break the $SO(3)$ symmetry, which has led to equal v.e.v's in the symmetric limit, and maintain the other symmetries of the Lagrangian. We will break the symmetry by  introducing symmetry breaking terms of dimension four. We can present a large number of symmetry breaking terms. The most straightforward way is to break the alignment of the v.e.v's of $(\epsilon_1,\; \epsilon_2)$ which, in turn, violate the decoupling in the neutrino mass matrix. Here, we introduce the most general form of symmetry breaking terms 
\bea
&&\xi \left( |\epsilon_1|^2 - |\epsilon_2|^2 \right)^2 + \left( |\epsilon_1|^2 - |\epsilon_2|^2 \right) \sum_{i=1}^2 \rho_i \phi_i^\dagger \phi_i + \varrho \left( |\epsilon_1|^4 - |\epsilon_2|^4 \right).
%
\label{sym-breaking}
\eea
The most general symmetry breaking terms can be expressed in terms of the form in Eq.~\ref{sym-breaking} and symmetry conserving terms that can be absorbed in the symmetric potential. Thus, the scalar potential including all the terms of the form in Eq.~\ref{sym-breaking} is given as follows
\bea
V&=& -\mu^2 \left( |\epsilon_1|^2 + |\epsilon_2|^2 + |\epsilon_3|^2 \right) + \left( |\epsilon_1|^2 + |\epsilon_2|^2 + |\epsilon_3|^2 \right) \sum_{i=1}^2 \sigma_i \phi_i^\dagger \phi_i   \nonumber\\
&+& \xi \left( |\epsilon_1|^2 - |\epsilon_2|^2 \right)^2 + \left( |\epsilon_1|^2 - |\epsilon_2|^2 \right) \sum_{i=1}^2 \rho_i \phi_i^\dagger \phi_i + \varrho \left( |\epsilon_1|^4 - |\epsilon_2|^4 \right) \nonumber\\
&+&\xi^\prime \left( |\epsilon_1|^2 + |\epsilon_2|^2 \right)^2 + \left( |\epsilon_1|^2 + |\epsilon_2|^2 \right) \sum_{i=1}^2 \rho^\prime_i \phi_i^\dagger \phi_i + \varrho^\prime \left( |\epsilon_1|^4 + |\epsilon_2|^4 \right)\nonumber\\
&+& \lambda \left( |\epsilon_1|^2 + |\epsilon_2|^2 + |\epsilon_3|^2 \right)^2 +V_{2HD}(\phi_1,\; \phi_2).
\label{potential6}
\eea
We can parametrize the v.e.v's of the singlet scalars as
\bea
\left\langle 0\left|\epsilon_1\right|0\right\rangle=\beta_1 \cos \gamma , \quad \left\langle 0\left|\epsilon_2\right|0\right\rangle=\beta_1 \sin \gamma,\;\mbox{and} \quad \left\langle 0\left|\epsilon_3\right|0\right\rangle = \beta_2 .
\label{beta1beta2}
\eea
We require that all terms in the symmetry breaking potential are of the same size which results in, from Eq.~\ref{sym-breaking}, $\varrho \sim \frac{v^2}{\beta_1^2}\rho_i$ and $\xi\sim \frac{v^2}{\beta_1^2 \cos 2\gamma}\rho_i$ where $v^2=v_1^2 + v_2^2$ is the EW scale. The only terms that depend on $ \gamma $ are
\bea
f(\gamma)=\xi \beta_1^4 \cos^2 2\gamma + \beta_1^2 \cos 2\gamma \sum_{i=1}^2 \rho_i |v_i|^2 + \varrho \beta_1^4 \cos 2\gamma +  \varrho^\prime \beta_1^4 \left( \frac{1+\cos^2 2\gamma}{2}\right).
\eea

After minimizing the potential, one can get the parameters of the v.e.v's as follows
\bea
\cos 2\gamma &=& -\frac{\varrho \beta_1^2+(\rho_1 |v_1|^2 + \rho_2 |v_2|^2 )}
{(2 \xi + \varrho^\prime) \beta_1^2}, \nonumber\\
\beta_1^2 &=&   \frac{|v_1|^2 (\varrho \rho_1 - \rho_1^\prime (2 \xi + \varrho^\prime)) + 
 |v_2|^2 (\varrho \rho_2 - \rho_2^\prime (2 \xi + \varrho^\prime))}{-\varrho^2 + 
  2 \xi (2 \xi^\prime + \varrho^\prime) + \varrho^\prime (2 \xi^\prime +  \varrho^\prime)}, \nonumber\\
  \beta_2^2 &=&   \frac{\beta_2^{\prime 2}}{-2\lambda(-\varrho^2 + 
    2 \xi (2 \xi^\prime + \varrho^\prime) + \varrho^\prime (2 \xi^\prime +  \varrho^\prime))},
\label{beta}
\eea
where 
\bea
\beta_2^{\prime 2} &\equiv& -\mu^2 (-\varrho^2 +2\xi (2\xi^\prime + \varrho^\prime) + \varrho^\prime (2 \xi^\prime +\varrho^{\prime }))\nonumber\\
&+& |v_1|^2 (\sigma_1(-\varrho^2 +2\xi (2\xi^\prime + \varrho^\prime) + \varrho^\prime (2 \xi^\prime +\varrho^{\prime })) + 2\lambda(\varrho \rho_1 - \rho_1^\prime (2 \xi + \varrho^\prime))) \nonumber\\
&+&|v_2|^2 (\sigma_2(-\varrho^2 +2\xi (2\xi^\prime + \varrho^\prime) + \varrho^\prime (2 \xi^\prime +\varrho^{\prime })) + 2\lambda(\varrho \rho_2 - \rho_2^\prime (2 \xi + \varrho^\prime))).
\eea
Then, we find that the following relation is satisfied
\beq
\beta_2^2 + \beta_1^2 = 3 w^2 .
\eeq
%
In Eq.~\ref{beta}, since $\varrho \sim \frac{v^2}{\beta_1^2}\rho_i$ that leads to $\cos 2\gamma\approx 0$ up to corrections of $v^2/\beta_1^2$ where $v$ is the EW scale and we assume $\beta_1$ to be in the TeV range in order to produce a sizable symmetry breaking parameter. 
However, we consider the first order correction to $\cos 2\gamma$ $(\cos 2\gamma \approx \tau)$ in our analysis where the symmetry breaking term is defined by
\beq
\tau \equiv -\frac{\varrho \beta_1^2+(\rho_1 |v_1|^2 + \rho_2 |v_2|^2 )}
{(2 \xi + \varrho^\prime) \beta_1^2}.
\label{cosine2}
\eeq
This leads to shifting the v.e.v's of the two singlet scalars  ($\left\langle 0\left|\epsilon_1\right|0\right\rangle\neq\left\langle 0\left|\epsilon_2\right|0\right\rangle$) up to the first order of $\tau$. Then, the Majorana neutrino mass matrix takes the form
\beq
M_R  =  \pmatrix{M+ v_{\beta_2} &  v_{\beta_{1 p}} &   v_{\beta_{1 n}}  \cr v_{\beta_{1 p}}   &  M   &  0 \cr   v_{\beta_{1 n}}  & 0  & M  } ,  \
\label{numatrixbreak}
\eeq
where $v_{\beta_i} = y \beta_i$ and
\bea
v_{\beta_{1 p}}&=&\frac{y}{\sqrt{2}}(\left\langle 0\left|\epsilon_1\right|0\right\rangle+\left\langle 0\left|\epsilon_2\right|0\right\rangle),\nonumber\\
v_{\beta_{1 n}}&=&\frac{y}{\sqrt{2}}(\left\langle 0\left|\epsilon_1\right|0\right\rangle-\left\langle 0\left|\epsilon_2\right|0\right\rangle). 
\label{pn}
\eea
We  write the v.e.v's of the singlet scalars after the symmetry breaking
 as  
\bea
\left\langle 0\left|\epsilon_1\right|0\right\rangle &= & \frac{\beta_1}{\sqrt{2}}\left( 1+\frac{\tau}{2}\right) ,\nonumber \\
\left\langle 0\left|\epsilon_2\right|0\right\rangle & = & \frac{\beta_1}{\sqrt{2}}\left( 1-\frac{\tau}{2}\right) ,
\eea
then
\bea
v_{\beta p}&=&v_{\beta_1},\nonumber\\
v_{\beta n}&=&\frac{\tau}{2} v_{\beta_1} . 
\label{pn2}
\eea
Note that, from the discussion below Eq.~\ref{beta1beta2} and using Eq.~\ref{cosine2} one finds
\beq
\xi\sim \frac{v^2}{\beta_1^2 \tau}\rho_i \sim \frac{\varrho}{\tau} ,
\eeq
which leads to $\xi \simeq 10 \varrho$ for $\tau=0.1$.

The results of the model have to satisfy the neutrino oscillation measurements. Although the numerical results show that breaking the $SO(3)$ symmetry in the scalar potential, see Eq.~\ref{sym-breaking}, is not sufficient to break the \textit{slight} equality of $(m_1,\; m_2)$ to satisfy the $\Delta m_{12}^2$ measurement. Therefore, we introduce additional terms to the Dirac mass term for the neutrinos which \textit{minimally} break the $SO(3)$ symmetry,
\beq
\frac{1}{2}\left[  M_1 \bar{\nu}_{e R} \nu_{e R}^c + M_2 \left( \bar{\nu}_{\mu R} \nu_{\mu R}^c + \bar{\nu}_{\tau R} \nu_{\tau R}^c \right) \right]  +h.c.
\label{Rashed-13}
\eeq
By presenting the above terms we have broken the $SO(3)$ symmetry in the whole Yukawa Lagrangian and the scalar potential. Note that the above terms break the $U(1)$ symmetry too. Thus
\beq
M_R  =  \pmatrix{M'+v_{\beta_2} & v_{\beta_1} &  \frac{\tau}{2} v_{\beta_1}  \cr v_{\beta_1}   &  M''   &  0 \cr  \frac{\tau}{2} v_{\beta_1}  & 0  & M''  },  \
\label{numatrixbreak}
\eeq
where $M'=M+M_1$ and $M''=M+M_2$. Using the see-saw formula (\ref{seesaw}), the neutrino mass matrix is given by
\bea
{\cal M}_\nu & = &  \pmatrix{X' & G' & P' \cr G' & Y' & W' \cr P' & W' & Z'},
\label{mass-seesaw-break}
\eea
where 
\bea
X' & = & -\frac{4 A^2 M''}{4M'M'' +4M'' v_{\beta_2} -v_{\beta_1}^2 (4+\tau^2)}  ,   \nonumber \\
Y' & = & -\frac{A^2 (4M'M''+4 M'' v_{\beta_2} -v_{\beta_1}^2 \tau^2)}{M'' (4M'M'' +4M'' v_{\beta_2} -v_{\beta_1}^2 (4+\tau^2))}   ,   \nonumber \\
Z' & = & -\frac{4 A^2 (M'M''+ M'' v_{\beta_2} - v_{\beta_1}^2 )}{M''(4M'M'' +4M'' v_{\beta_2} -v_{\beta_1}^2 (4+\tau^2))}  ,   \nonumber \\
G' & = & \frac{4 A^2 v_{\beta_1}}{4M'M'' +4M'' v_{\beta_2} -v_{\beta_1}^2 (4+\tau^2)} , \nonumber \\
P' & = & \frac{2 A^2  v_{\beta_1} \tau}{4M'M'' +4M'' v_{\beta_2} -v_{\beta_1}^2 (4+\tau^2)},\nonumber \\
W' & = & -\frac{2 A^2 v_{\beta_1}^2 \tau}{M'' (4M'M'' +4M'' v_{\beta_2} -v_{\beta_1}^2 (4+\tau^2))}  .   
\label{mass-seesaw-break2}
\eea
From Eqs.~(\ref{mass-seesaw-break}, \ref{mass-seesaw-break2}), one gets the mass eigenvalues
\begin{eqnarray}
m_1 &=&-   \frac{A^2}{4M'M'' + 4 M'' v_{\beta_2} -v_{\beta_1}^2 (4+\tau^2)}\left[ 2(M'+M''+v_{\beta_2})\right. \nonumber\\
&&\left. - 2\sqrt{M'^2 +M''^2 -2M'(M''-v_{\beta_2})-2M'' v_{\beta_2} +v_{\beta_2}^2+v_{\beta_1}^2 (4+\tau^2)}\right], \nonumber\\
m_2 &=&-   \frac{A^2}{4M'M'' + 4 M'' v_{\beta_2} -v_{\beta_1}^2 (4+\tau^2)}\left[ 2(M'+M''+v_{\beta_2})\right. \nonumber\\
    & &\left.  + 2\sqrt{M'^2 +M''^2 -2M'(M''-v_{\beta_2})-2M'' v_{\beta_2} +v_{\beta_2}^2+v_{\beta_1}^2 (4+\tau^2)}\right] ,\nonumber\\
m_3 &=&-   \frac{A^2}{M''}.
\label{masses}
\end{eqnarray}
We can diagonalize the mass matrix in Eq.~\ref{mass-seesaw-break} using the unitary matrix
 $U_{\nu}=W^{\nu}_{12}R_{23}^{\nu}R_{12}^{\nu}$ with,
\bea
 R_{12}^{\nu} & = & \pmatrix{ c_{12\nu} & s_{12\nu} & 0 \cr 
                              -s_{12\nu} & c_{12\nu} & 0 \cr
                              0 & 0 & 1 }, \nonumber\\
c_{12\nu} & = & \cos { \theta_{12\nu}} ; s_{12\nu}  =  \sin { \theta_{12\nu}},\nonumber\\
R_{23}^{\nu} & = & \pmatrix{ 1 & 0 & 0 \cr 
                              0 & c_{23\nu} & s_{23\nu} \cr
                              0 & -s_{23\nu} & c_{23\nu} }, \nonumber\\
c_{23\nu} & = & \cos { \theta_{23\nu}} ; s_{23\nu}  =  \sin { \theta_{23\nu}}.
\eea
On can find relations between the mass matrix elements in Eq.~\ref{mass-seesaw-break2}
\begin{eqnarray}
 X'(Z'-Y')&=&P'^2-G'^2, \nonumber\\
 G' P' (Z'-Y')&=& W'(P'^2-G'^2).
\end{eqnarray}
Applying the above relations to the corresponding mass matrix elements of ${\cal {M}}_\nu = U_\nu {\cal {M}}_\nu^d U_\nu^\dagger $ with $U_{\nu}=W^{\nu}_{12}R_{23}^{\nu}R_{12}^{\nu}$ , one can get the two mixing angles
\bea
s_{23\nu} &=&  \sqrt{\frac{3  m_1 (m_2 - m_3)}{m_2 (m_1 - m_3)}}, \nonumber\\
s_{12\nu} &=&\sqrt{\frac{-2 m_1 m_2 + 3 m_1 m_3 - m_2 m_3 }{3 m_3 (m_1 - m_2) }} .
\label{nu-correction-angles}
\eea

Following Ref.~\cite{king}, we expand the angles in Eq.~\ref{standpara} as
\begin{equation}
s_{13} = \frac{r}{\sqrt{2}}, \ \ s_{12} = \frac{1}{\sqrt{3}}(1+s), \ \ s_{23} = \frac{1}{\sqrt{2}}(1+a),
\label{rsa}
\end{equation}
where the three real parameters
$r,s,a$  describe the deviations of the reactor, solar, and
atmospheric angles, respectively, from their tri-bimaximal values. We  use
global fits of the mixing parameters with $3 \sigma$ significance \cite{Tortola:2012te}
\begin{eqnarray}
0.18<r<0.26, \ -0.10<s<0.05, \ -0.15<a<0.17. 
\label{ranges}
\end{eqnarray}
To first order in $r,s,a$ the lepton mixing matrix can be written as \cite{king},
\begin{eqnarray}
U \approx
\left( \begin{array}{ccc}
\sqrt{\frac{2}{3}}(1-\frac{1}{2}s)  & \frac{1}{\sqrt{3}}(1+s) & \frac{1}{\sqrt{2}}re^{-i\delta } \\
-\frac{1}{\sqrt{6}}(1+s-a + re^{i\delta })  & \frac{1}{\sqrt{3}}(1-\frac{1}{2}s-a- \frac{1}{2}re^{i\delta })
& \frac{1}{\sqrt{2}}(1+a) \\
\frac{1}{\sqrt{6}}(1+s+a- re^{i\delta })  & -\frac{1}{\sqrt{3}}(1-\frac{1}{2}s+a+ \frac{1}{2}re^{i\delta })
 & \frac{1}{\sqrt{2}}(1-a)
\end{array}
\right).
\label{MNS1}
\end{eqnarray}
We are not going to consider CP violation in this work, thus, we assume that $\delta=0$. We can write the parameters $(r,\; s,\; a)$ in terms of the elements of the mixing matrix, 
\bea
r & = & -1-s+a-\sqrt{6} U_{21}, \nonumber\\
s & = & -1+\sqrt{3} U_{12}, \nonumber\\
a & = & -1+\sqrt{2} U_{23}.
\label{Full-Equations2}
\eea 
Now, we can calculate the full deviation of the leptonic mixing coming from the charged lepton and neutrino sector. We obtain the elements of the lepton mixing matrix
\beq
U_{PMNS}=U_{l}^{\dag}U_{\nu},
\label{Upmns}
\eeq
with $U_\ell  =  W^l_{23} R^l_{23} R^l_{13} R^l_{12}$ and $U_{\nu}=W^{\nu}_{12}R_{23}^{\nu}R_{12}^{\nu}$. Thus, up to first order in $(s_{12l},s_{13l},s_{23l})$ one can get
\begin{eqnarray}
 r & \approx &  -s_{12l}+\sqrt{\frac{2}{3}}s_{23\nu}+s_{13l}, \nonumber\\
 s & \approx &  -s_{12l}+\sqrt{2}s_{12\nu}-s_{13l}, \nonumber\\
 a & \approx &  -s_{23l}+\sqrt{\frac{2}{3}}s_{23\nu}. \
 \label{final}
\end{eqnarray}

In Ref.~\cite{Rashed:2011zs}, it was found that the contribution of the charged lepton sector, with $\delta=0$, is give as 
\begin{itemize}
	\item For $z=1.8$: $s_{12l}\approx\pm 0.44,  \;s_{13l}\approx\mp 0.0012,\; s_{23l}\approx - 0.053$,
	\item For $z=1.7$: $s_{12l}\approx\pm 0.48,  \;s_{13l}\approx\mp 0.0013,\; s_{23l}\approx - 0.050$,
\end{itemize}
We can check the contributions of the charged leptons, $U_\ell  =  W^l_{23} R^l_{23} R^l_{13} R^l_{12}$, without corrections from the neutrino sector, i.e. $U_{\nu}=W^{\nu}_{12}$. By substituting the above values in Eq.~\ref{final}, up to the first order one gets
\begin{itemize}
	\item For $z=1.8$:  $ r \approx 0.44,\; s \approx 0.44,\; a \approx 0.053$,
	\item For $z=1.7$:  $ r \approx 0.48,\; s \approx 0.48,\; a \approx 0.050$,
\label{numbericalresults2}
\end{itemize}
The results above do not match the experimental values where the charged lepton sector introduces  large corrections to the mixing angles $\theta_{13}$ and $\theta_{12}$. Thus, it becomes necessary  to combine the contributions come from the charged lepton and neutrino sector in order to calculate the full deviation from the TBM mixing.


\section{Numerical results}


In the case of degenerate neutrino masses $m_1\approx m_2\approx m_3$, one can find from Eq.~\ref{mass-eigenvalues} that $a\approx b \approx c$. This leads to a diagonal neutrino mass matrix ${\cal {M}}_{\nu} \approx \mbox{diag}\left(a,a,a\right)$. This means that the lepton mixing matrix does not include a contribution from the neutrino sector, which is inconsistent with the experimental data. Thus, in the symmetric limit our model excludes the  case of the degenerate neutrino masses.

\begin{figure}[h!]
	\centering
		\includegraphics[width=7.4cm]{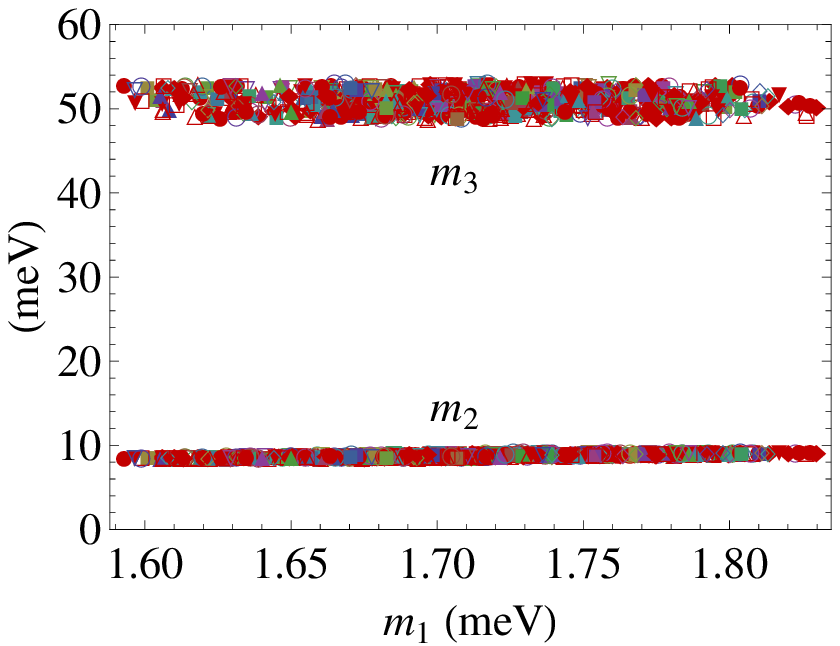}
		\includegraphics[width=7.4cm]{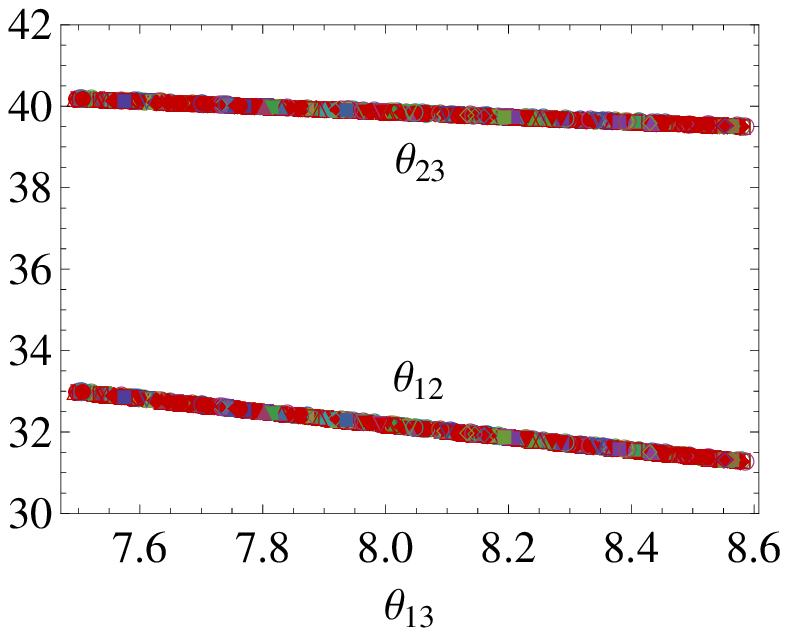}
		\includegraphics[width=7.4cm]{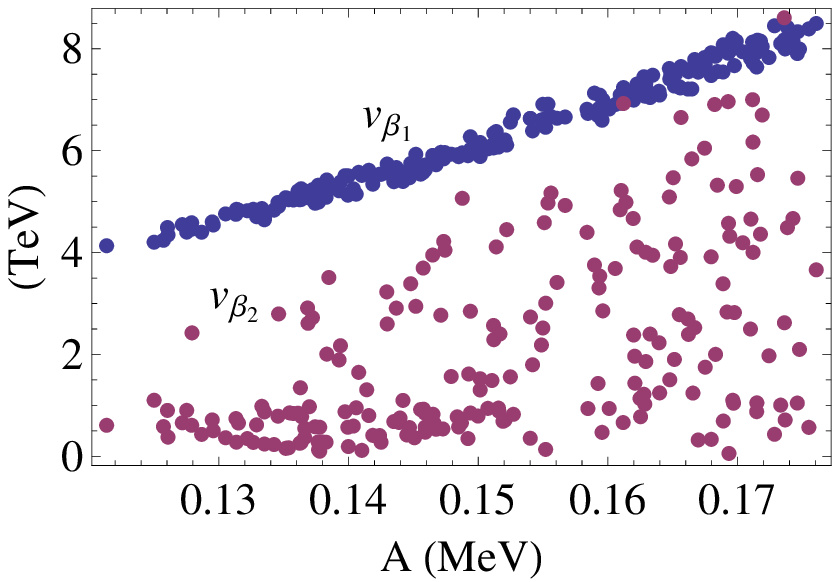}
		\includegraphics[width=7.4cm]{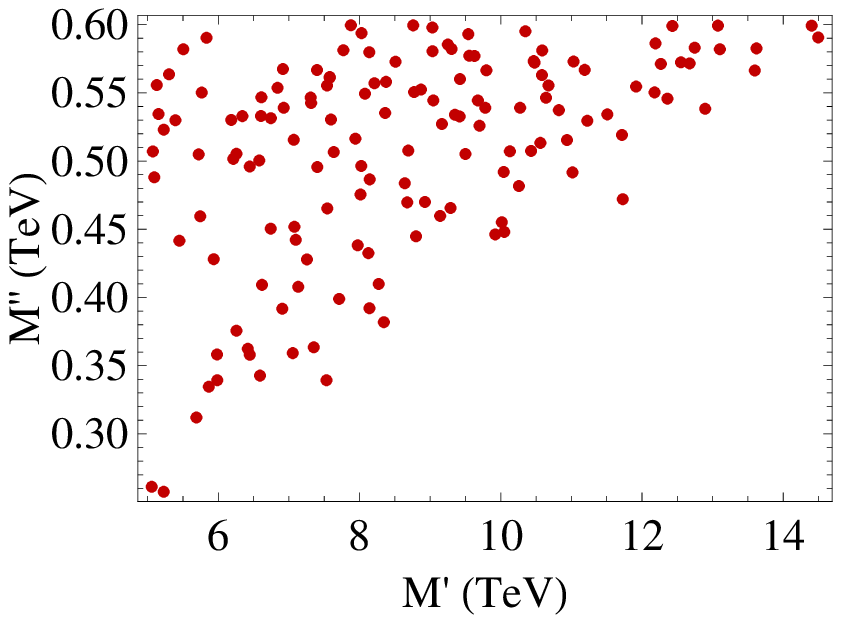}
		\includegraphics[width=7.4cm]{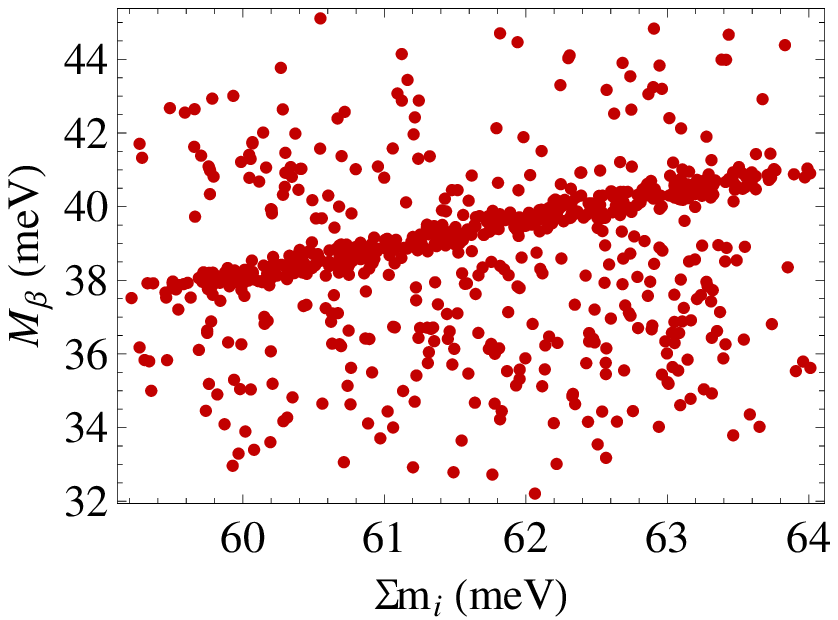}
		\includegraphics[width=7.4cm]{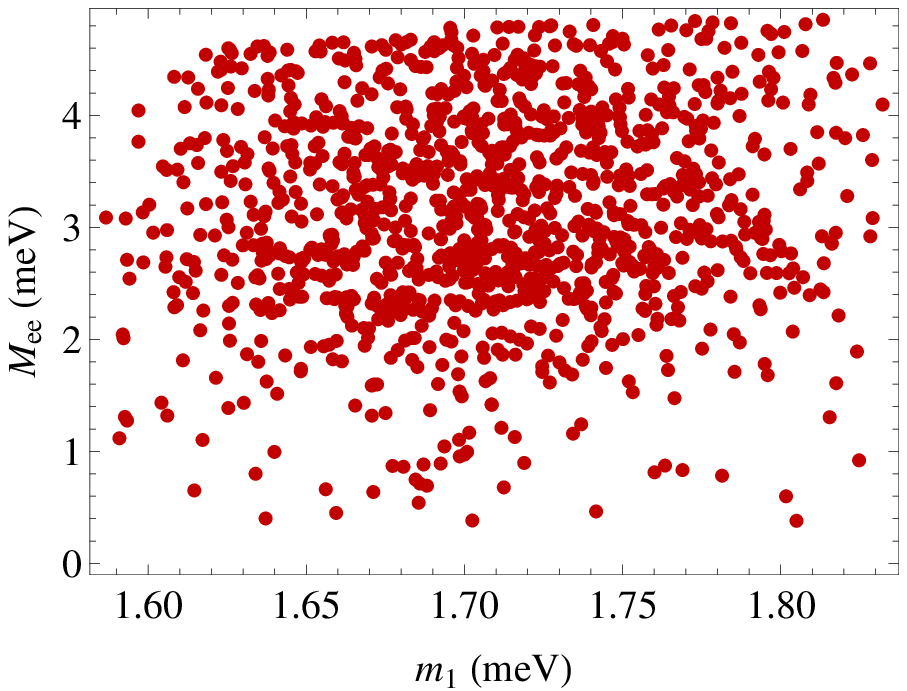}
	\caption{Scatter plot for $z=1.8$ with $s_{12l}\approx  -0.44,\; s_{13l}\approx  0.0012$, and  $s_{23l} \approx -0.053$. In the neutrino sector, we take $\tau=0.1 $ (meV $= 10^{-3}$ eV). In the $M_{ee}$-$m_{1}$ graph, the Majorana phases $(\zeta_1 ,\; \zeta_2)$ are varied in the interval $[0,\pi]$.}
	\label{plot1}
\end{figure}

\begin{figure}[h!]
	\centering
		\includegraphics[width=7.4cm]{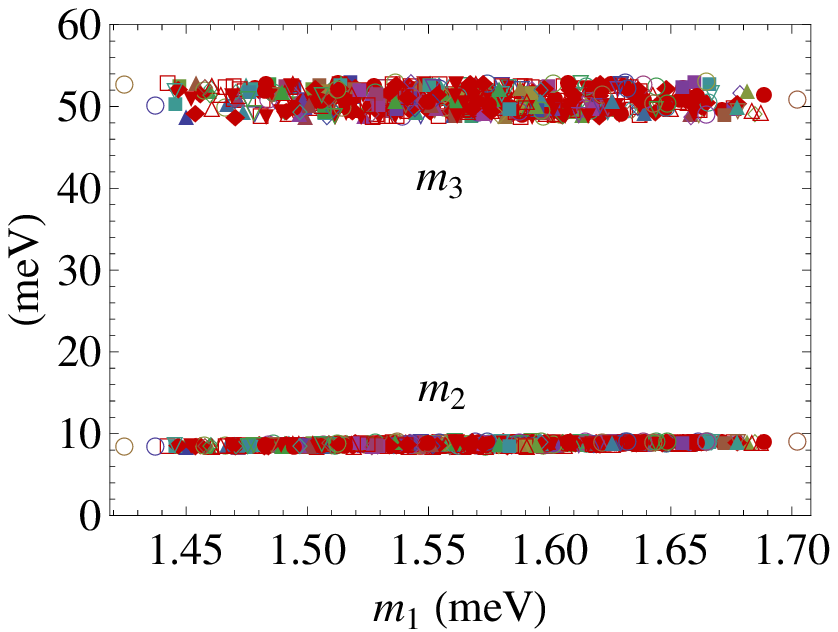}
		\includegraphics[width=7.4cm]{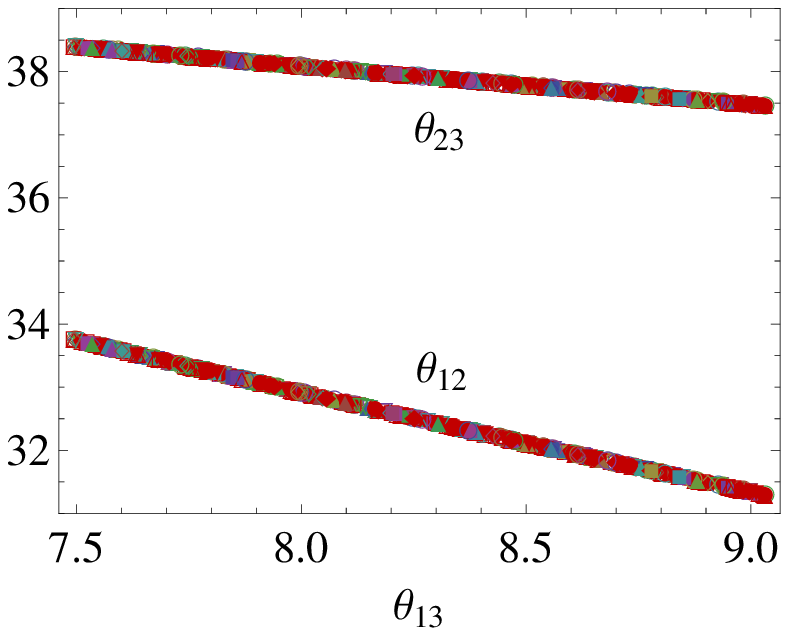}
		\includegraphics[width=7.4cm]{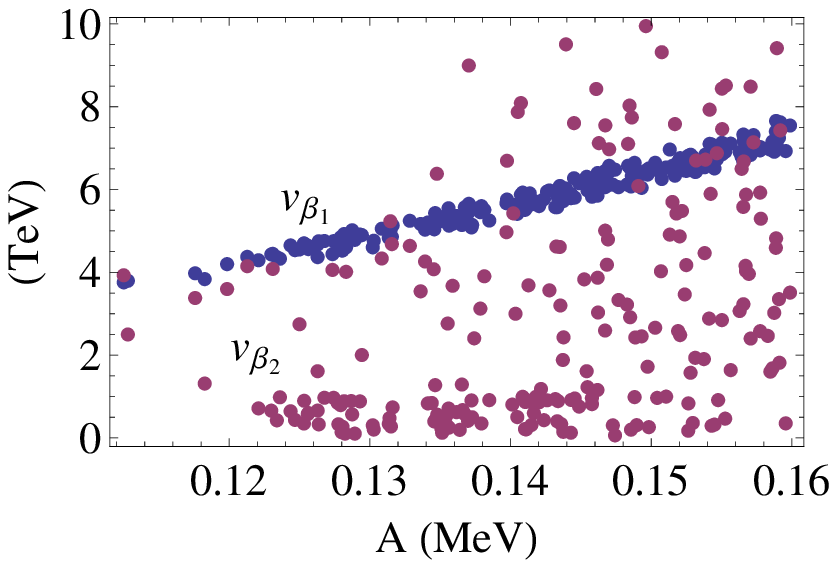}
		\includegraphics[width=7.4cm]{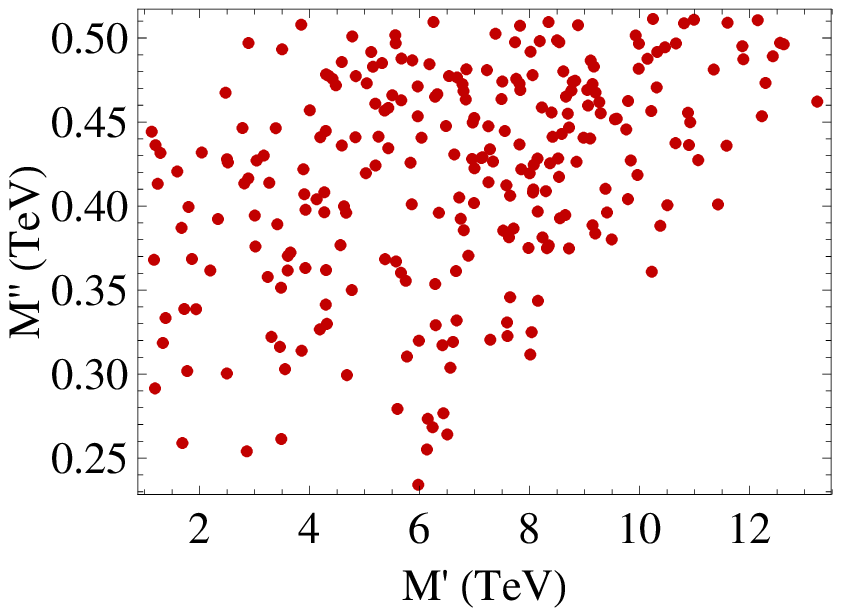}
		\includegraphics[width=7.4cm]{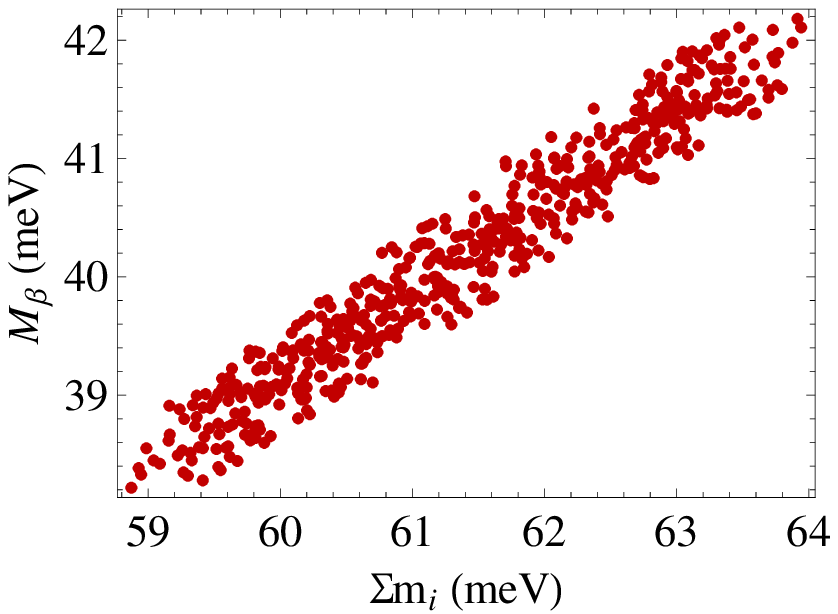}
		\includegraphics[width=7.4cm]{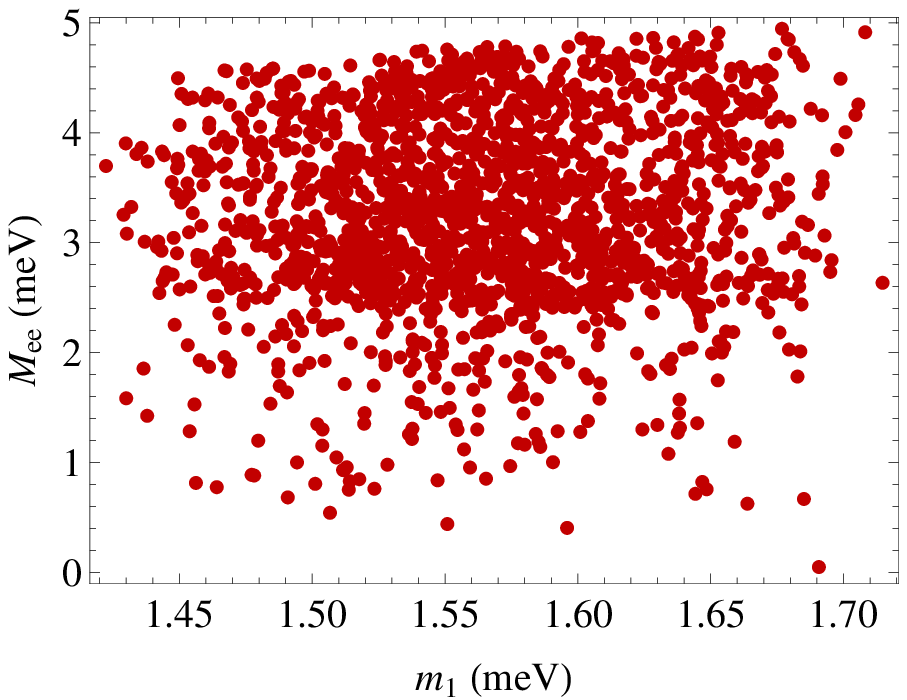}
	\caption{Scatter plot for $z=1.7$ with $s_{12l}\approx  -0.48,\; s_{13l}\approx  0.0013$, and  $s_{23l} \approx -0.05$. In the neutrino sector, we take $\tau=0.05 $ (meV $= 10^{-3}$ eV). In the $M_{ee}$-$m_{1}$ graph, the Majorana phases $(\zeta_1 ,\; \zeta_2)$ are varied in the interval $[0,\pi]$.}
	\label{plot2}
\end{figure}

The numerics goes as follows; using the experimental values of the neutrino mixing parameters at $3\sigma$ significance \cite{Tortola:2012te} we choose random values of the masses $(m_1 , m_2 , m_3)$ which satisfy
\bea
\Delta m_{21}^2 &=& m_2^2 - m_1^2 = (7.12-8.20)\times 10^{-5} eV^2,\nonumber\\
\Delta m_{32}^2 &=& |m_3^2 - m_2^2| = (2.31-2.74)\times 10^{-3} eV^2. 
\eea
We substitute the masses in ($r, s, a$) in Eq.~\ref{Full-Equations2} with $(s_{12\nu},\; s_{23\nu})$  given in Eq.~\ref{nu-correction-angles} and $(s_{12l},\; s_{23l},\; s_{13l})$  in the previous section. If the results agree with the experimental constraints in Eq.~\ref{ranges} we plot, in Figs.~(\ref{plot1}, \ref{plot2}), the possible values of the absolute masses and  mixing angles. By substituting the masses obtained above in Eq.~\ref{masses}, one can find the values of the Lagrangian parameters $(v_{\beta 1},\; v_{\beta 2},\; A,\; M',\; M'')$. The results support the normal mass hierarchy. The figures show that the scale of the neutrino masses is in the few meV to $ \sim$ 50 meV range (meV$=10^{-3}$eV). Also, the  full contribution from both the charged lepton and neutrino sector accommodates the measurements of the mixing angles. The graphs (\ref{plot1}, \ref{plot2}) show that the see-saw scales $(M',\; M'')$ are in the TeV range, and the extra Higgs, that generates the Dirac neutrino masses, has v.e.v $(v_2)$, included in $A$, in the MeV scale{\footnote{Higgs doublet with a small v.e.v has been discussed in the literatures \cite{smallVEV}.}}. Also, the graphs indicate that the v.e.v's of the singlet scalar fields $(v_{\beta 1},\; v_{\beta 2})$ are in the TeV scale. {\footnote{Several papers have introduced neutrino mixing models in the TeV scale (for review see Refs.~\cite{TeV-1,TeV-2}).}} Various other mechanisms to generate the neutrino masses with TeV scale new physics are mentioned in Ref.~\cite{Chen:2011de}.

 Three mass-dependent neutrino observables are probed in different types of experiments. The sum of absolute neutrino masses $m_{cosm}\equiv\Sigma m_i$ is probed in cosmology, the kinetic electron neutrino mass in the beta decay $(M_\beta)$ is probed in direct search for neutrino masses, and the effective mass $(M_{ee} )$ in neutrinoless double beta decay $0\nu\beta\beta$ is probed in the $0\nu\beta\beta$ experiment where the decay rate $\Gamma \propto M_{ee}^2$. In terms of the ``bare'' physical parameters, $m_i$ and $U_{\alpha i}$, the observables are given by \cite{Barry:2010yk}
 \bea
 \Sigma m_i &=& |m_1|+|m_2|+|m_3|, \nonumber\\
 M_{ee}  &=& ||m_1||U_{e1}|^2 +|m_2||U_{e2}|^2 e^{i\zeta_1}+|m_3||U_{e3}|^2 e^{i\zeta_2}|, \nonumber\\
 M_\beta &=& \sqrt{|m_1|^2 |U_{e1}|^2 +|m_2|^2 |U_{e2}|^2 +|m_3|^2 |U_{e3}|^2},
 \eea
We plot $M_\beta$ versus $\Sigma m_i$ and $M_{ee}$ versus $m_{light}$, where $m_{light}$ is the lightest neutrino mass which is $m_1$ in this model and the Majorana phases $(\zeta_1 ,\; \zeta_2)$ are varied in the interval $[0,\pi]$. The graphs show that $\Sigma m_i\approx 60$ meV, $ M_{ee} < M_\beta $, $M_{ee} < 0.40$ eV \cite{Mohapatra:2006gs}, and the graphical representation $M_{ee}$-$m_{1}$ agrees with the results in Ref.~\cite{Beringer:1900zz}.



\section{ Conclusion}


In this paper, we extended our model in Ref.~\cite{Rashed:2011zs} to treat the leptonic mixing in the flavor symmetric limit with the tri-bimaximal pattern. The charged lepton sector was considered in Ref.~\cite{Rashed:2011zs} in a basis where the charged lepton Yukawa matrix is non-diagonal with a $2-3$ symmetric structure except for one breaking by the muon mass. We fixed the neutrino mass matrix to have a decoupling of the first two generations from the third one, and under a certain condition we generated the lepton mixing in the symmetric limit with the TBM structure. This model was described by the Lagrangian that extended the SM by three right-handed neutrinos, an extra Higgs doublet, and three complex singlet scalar fields. Also, the symmetry group of the SM was extended by the product of the symmetries $Z_4\times U(1)\times SO(3)$. The global $SO(3)$ flavour symmetry was introduced in a way that it is maintained by the Dirac neutrino mass terms in the Yukawa Lagrangian and the scalar potential and broken elsewhere.

The symmetry breaking in the charged lepton sector did not fit the data by introducing a large contribution to the mixing angles $\theta_{13}$ and $\theta_{12}$. Therefore, by breaking the $SO(3)$ symmetry in the effective potential in a way that violates the alignment of the v.e.v's of the singlet scalars and in the entire Yukawa Lagrangian, the contribution of the neutrino sector to the symmetry breaking was introduced to accommodate the measurements. The analysis of our model to fit the experimental constraints of the mixing angles showed that this model supported the normal mass hierarchy with masses in the few meV to $\sim 50$ meV range, meV$=10^{-3}$eV. Also, the v.e.v of the additional Higgs was obtained in the MeV scale. The v.e.v's of the singlet scalars and the see-saw scale were found to be in the TeV range. The graphs showed that $\Sigma m_i\approx 60$ meV and $ M_{ee} < M_\beta $ and $M_{ee} < 0.40$ eV. The graphical representation $M_{ee}$-$m_{1}$ agreed with the results in Ref.~\cite{Beringer:1900zz}.


\section{Acknowledgements}


The author gratefully thanks A.~Datta for enlightening discussions and comments. Also, the author thanks Yue-Liang Wu for suggesting useful references. This work was supported in part by the US-Egypt Joint Board on Scientific and Technological Co-operation award (Project ID: 1855) administered by the US Department of Agriculture, summer grant from the College of Liberal Arts, University of Mississippi, and in part by the National Science Foundation under Grant No. 1068052.

\pagebreak



\pagebreak

\end{document}